
\magnification=1200
\baselineskip=14truept
\font\seven=cmr7

\font\nine=cmr9
\font\bcp=cmbx7

\font\bs=cmbx12 
\font\text=cmr10
\font\it=cmti10
\def\pmb#1{\setbox0=\hbox{#1}%
  \hbox{\kern-.025em\copy0\kern-\wd0
  \kern.05em\copy0\kern-\wd0
  \kern-0.025em\raise.0433em\box0} }

\def\half {{1\over 2}}

\catcode`@=11
\def\leftrightarrowfill{$\m@th\mathord\leftarrow \mkern-6mu
  \cleaders\hbox{$\mkern-2mu \mathord- \mkern-2mu$}\hfill
  \mkern-6mu \mathord\rightarrow$}
\def\overleftrightarrow#1{\vbox{\ialign{##\crcr
     \leftrightarrowfill\crcr\noalign{\kern-1pt\nointerlineskip}
     $\hfil\displaystyle{#1}\hfil$\crcr}}}
\catcode`@=12

\def\approxge{\hbox {\hfil\raise .4ex\hbox{$>$}\kern-.75 em
\lower .7ex\hbox{$\sim$}\hfil}}
\def\approxle{\hbox {\hfil\raise .4ex\hbox{$<$}\kern-.75 em
\lower .7ex\hbox{$\sim$}\hfil}}

\def \abstract#1 {\vskip 0.5truecm\sepline\vskip 0.5truecm
$$\vbox{\hsize=15truecm\noindent #1}$$}
\def \SISSA#1#2 {\vfil\vfil\centerline{Ref. S.I.S.S.A. #1 CM (#2)}}
\def \PACS#1 {\vfil\line{\hfil\hbox to 15truecm{PACS numbers: #1 \hfil}\hfil}}
\def \hfigure
     #1#2#3       {\midinsert \vskip #2 truecm $$\vbox{\hsize=14.5truecm
             \seven\baselineskip=10pt\noindent {\bcp \noindent Figure  #1}.
                   #3 } $$ \vskip -20pt \endinsert }
\def \hfiglin
     #1#2#3       {\midinsert \vskip #2 truecm $$\vbox{\hsize=14.5truecm
              \seven\baselineskip=10pt\noindent {\bcp \hfil\noindent
                   Figure  #1}. #3 \hfil} $$ \vskip -20pt \endinsert }
\def \vfigure
     #1#2#3#4     {\dimen0=\hsize \advance\dimen0 by -#3truecm
                   \midinsert \vbox to #2truecm{ \seven
                   \parshape=1 #3truecm \dimen0 \baselineskip=10pt \vfill
                   \noindent{\bcp Figure #1} \pretolerance=6500#4 \vfill }
                   \endinsert }

\def \ref
     #1#2         {\smallskip \item{[#1]}#2}
\def \sepline     {\medskip\centerline{\vbox{\hrule width5truecm}} \medskip}
\def \tabrule     {\noalign{\vskip 5truept \hrule\vskip 5truept} }
\def \tabrul2     {\noalign{\vskip 5truept \hrule \vskip 2truept \hrule
                   \vskip 5truept} }
\footline={\ifnum\pageno>0 \tenrm \hss \folio \hss \fi }
\def\today
 {\count10=\year\advance\count10 by -1900 \number\day--\ifcase
  \month \or Jan\or Feb\or Mar\or Apr\or May\or Jun\or
             Jul\or Aug\or Sep\or Oct\or Nov\or Dec\fi--\number\count10}
\def\hour{\count10=\time\count11=\count10
\divide\count10 by 60 \count12=\count10
\multiply\count12 by 60 \advance\count11 by -\count12\count12=0
\number\count10 :\ifnum\count11 < 10 \number\count12\fi\number\count11}
\def\draft{
   \baselineskip=20pt
   \def\makeheadline{\vbox to 10pt{\vskip-22.5pt
   \line{\vbox to 8.5pt{}\the\headline}\vss}\nointerlineskip}
   \headline={\hfill \seven {\bcp Draft version}: today is \today\ at \hour
              \hfill}
          }
%
%
%
\catcode`@=11
%
%
\def\b@lank{ }
\newif\if@simboli
\newif\if@riferimenti
\newwrite\file@simboli
\def\simboli{
    \immediate\write16{ !!! Genera il file \jobname.SMB }
    \@simbolitrue\immediate\openout\file@simboli=\jobname.smb}
\newwrite\file@ausiliario
\def\riferimentifuturi{
    \immediate\write16{ !!! Genera il file \jobname.AUX }
    \@riferimentitrue\openin1 \jobname.aux
    \ifeof1\relax\else\closein1\relax\input\jobname.aux\fi
    \immediate\openout\file@ausiliario=\jobname.aux}
\newcount\eq@num\global\eq@num=0
\newcount\sect@num\global\sect@num=0
\newif\if@ndoppia
\def\numerazionedoppia{\@ndoppiatrue\gdef\la@sezionecorrente{\the\sect@num}}
\def\se@indefinito#1{\expandafter\ifx\csname#1\endcsname\relax}
\def\spo@glia#1>{} 
\newif\if@primasezione
\@primasezionetrue
\def\s@ection#1\par{\immediate
    \write16{#1}\if@primasezione\global\@primasezionefalse\else\goodbreak
    \vskip\spaziosoprasez\fi\noindent
    {\bf#1}\nobreak\vskip\spaziosottosez\nobreak\noindent}
%

\def\sezpreset#1{\global\sect@num=#1
    \immediate\write16{ !!! sez-preset = #1 }   }
\def\spaziosoprasez{50pt plus 60pt}
\def\spaziosottosez{15pt}
\def\sref#1{\se@indefinito{@s@#1}\immediate\write16{ ??? \string\sref{#1}
    non definita !!!}
    \expandafter\xdef\csname @s@#1\endcsname{??}\fi\csname @s@#1\endcsname}
\def\autosez#1#2\par{
    \global\advance\sect@num by 1\if@ndoppia\global\eq@num=0\fi
    \xdef\la@sezionecorrente{\the\sect@num}
    \def\usa@getta{1}\se@indefinito{@s@#1}\def\usa@getta{2}\fi
    \expandafter\ifx\csname @s@#1\endcsname\la@sezionecorrente\def
    \usa@getta{2}\fi
    \ifodd\usa@getta\immediate\write16
      { ??? possibili riferimenti errati a \string\sref{#1} !!!}\fi
    \expandafter\xdef\csname @s@#1\endcsname{\la@sezionecorrente}
    \immediate\write16{\la@sezionecorrente. #2}
    \if@simboli
      \immediate\write\file@simboli{ }\immediate\write\file@simboli{ }
      \immediate\write\file@simboli{  Sezione
                                  \la@sezionecorrente :   sref.   #1}
      \immediate\write\file@simboli{ } \fi
    \if@riferimenti
      \immediate\write\file@ausiliario{\string\expandafter\string\edef
      \string\csname\b@lank @s@#1\string\endcsname{\la@sezionecorrente}}\fi
    \goodbreak\vskip 48pt plus 60pt
    \noindent{\title\the\sect@num.\quad #2}\par\nobreak\vskip 15pt
    \nobreak\noindent}
\def\semiautosez#1#2\par{
    \gdef\la@sezionecorrente{#1}\if@ndoppia\global\eq@num=0\fi
    \if@simboli
      \immediate\write\file@simboli{ }\immediate\write\file@simboli{ }
      \immediate\write\file@simboli{  Sezione ** : sref.
          \expandafter\spo@glia\meaning\la@sezionecorrente}
      \immediate\write\file@simboli{ }\fi
    \s@ection#2\par}


\def\eqpreset#1{\global\eq@num=#1
     \immediate\write16{ !!! eq-preset = #1 }     }

\def\eqref#1{\se@indefinito{@eq@#1}
    \immediate\write16{ ??? \string\eqref{#1} non definita !!!}
    \expandafter\xdef\csname @eq@#1\endcsname{??}
    \fi\csname @eq@#1\endcsname}

\def\eqlabel#1{\global\advance\eq@num by 1
    \if@ndoppia\xdef\il@numero{\la@sezionecorrente.\the\eq@num}
       \else\xdef\il@numero{\the\eq@num}\fi
    \def\usa@getta{1}\se@indefinito{@eq@#1}\def\usa@getta{2}\fi
    \expandafter\ifx\csname @eq@#1\endcsname\il@numero\def\usa@getta{2}\fi
    \ifodd\usa@getta\immediate\write16
       { ??? possibili riferimenti errati a \string\eqref{#1} !!!}\fi
    \expandafter\xdef\csname @eq@#1\endcsname{\il@numero}
    \if@ndoppia
       \def\usa@getta{\expandafter\spo@glia\meaning
       \la@sezionecorrente.\the\eq@num}
       \else\def\usa@getta{\the\eq@num}\fi
    \if@simboli
       \immediate\write\file@simboli{  Equazione
            \usa@getta :  eqref.   #1}\fi
    \if@riferimenti
       \immediate\write\file@ausiliario{\string\expandafter\string\edef
       \string\csname\b@lank @eq@#1\string\endcsname{\usa@getta}}\fi}

\def\autoeqno#1{\eqlabel{#1}\eqno(\csname @eq@#1\endcsname)}
\def\autoleqno#1{\eqlabel{#1}\leqno(\csname @eq@#1\endcsname)}
\def\eqrefp#1{(\eqref{#1})}


\def\eq{\autoeqno}
\def\req{\eqrefp}



\newcount\cit@num\global\cit@num=0

\newwrite\file@bibliografia
\newif\if@bibliografia
\@bibliografiafalse

\def\lp@cite{[}
\def\rp@cite{]}
\def\trap@cite#1{\lp@cite #1\rp@cite}
\def\lp@bibl{\ \ }
\def\rp@bibl{.}
\def\trap@bibl#1{\lp@bibl #1\rp@bibl}

\def\refe@renza#1{\if@bibliografia\immediate        
    \write\file@bibliografia{
    \string\item{\trap@bibl{\cref{#1}}}\string
    \bibl@ref{#1}\string\bibl@skip}\fi}

\def\ref@ridefinita#1{\if@bibliografia\immediate\write\file@bibliografia{
    \string\item{?? \trap@bibl{\cref{#1}}} ??? tentativo di ridefinire la
      citazione #1 !!! \string\bibl@skip}\fi}

\def\bibl@ref#1{\se@indefinito{@ref@#1}\immediate
    \write16{ ??? biblitem #1 indefinito !!!}\expandafter\xdef
    \csname @ref@#1\endcsname{ ??}\fi\csname @ref@#1\endcsname}

\def\c@label#1{\global\advance\cit@num by 1\xdef            
   \la@citazione{\the\cit@num}\expandafter
   \xdef\csname @c@#1\endcsname{\la@citazione}}

\def\bibl@skip{\vskip +4truept}


\def\stileincite#1#2{\global\def\lp@cite{#1}\global
    \def\rp@cite{#2}}
\def\stileinbibl#1#2{\global\def\lp@bibl{#1}\global
    \def\rp@bibl{#2}}

\def\citpreset#1{\global\cit@num=#1
    \immediate\write16{ !!! cit-preset = #1 }    }

\def\autobibliografia{\global\@bibliografiatrue\immediate
    \write16{ !!! Genera il file \jobname.BIB}\immediate
    \openout\file@bibliografia=\jobname.bib}

\def\cref#1{\se@indefinito                  
   {@c@#1}\c@label{#1}\refe@renza{#1}\fi\csname @c@#1\endcsname}

\def\cite#1{\trap@cite{\cref{#1}}}                  
\def\ccite#1#2{\trap@cite{\cref{#1},\cref{#2}}}     
\def\cccite#1#2#3{\trap@cite{\cref{#1},\cref{#2},\cref{#3}}}     
\def\ncite#1#2{\trap@cite{\cref{#1}--\cref{#2}}}    
\def\upcite#1{$^{\,\trap@cite{\cref{#1}}}$}               
\def\upccite#1#2{$^{\,\trap@cite{\cref{#1},\cref{#2}}}$}  
\def\upncite#1#2{$^{\,\trap@cite{\cref{#1}-\cref{#2}}}$}  

\def\clabel#1{\se@indefinito{@c@#1}\c@label           
    {#1}\refe@renza{#1}\else\c@label{#1}\ref@ridefinita{#1}\fi}

\def\biblskip#1{\def\bibl@skip{\vskip #1}}           

\def\insertbibliografia{\if@bibliografia             
    \immediate\write\file@bibliografia{ }
    \immediate\closeout\file@bibliografia
    \catcode`@=11\input\jobname.bib\catcode`@=12\fi}


\def\commento#1{\relax}
\def\biblitem#1#2\par{\expandafter\xdef\csname @ref@#1\endcsname{#2}}


\catcode`@=12



\def\interlinea{\baselineskip=14truept}
\interlinea

\global\newcount\nnote \global\nnote=0
\def\note #1 {\global\advance\nnote by1 \baselineskip 10pt
              \footnote{$^{\the\nnote}$}{\nine #1} \interlinea}

\autobibliografia\stileincite{}{}

\def\date{\hfill Trieste,\space\ifcase\month\or
 January\or February\or March\or April\or May\or June\or
 July\or August\or September\or October\or November\or December\fi
\space\number\day ,\space \number\year}

\def\npb{Nucl. Phys. B}
\def\prd{Phys. Rev. D}
\def\plb{Phys. Lett. B}
\def\zpc{Z. Phys. C}
\def\prl{Phys. Rev. Lett. }

\def\hb{\hfil\break}
\def\no{\noindent}
\def\bs{\bigskip}

\def\gtrsim{\ \rlap{\raise 2pt \hbox{$>$}}{\lower 2pt \hbox{$\sim$}}\ }
\def\lesssim{\ \rlap{\raise 2pt \hbox{$<$}}{\lower 2pt \hbox{$\sim$}}\ }

\def\half{{1\over 2}}
\def\n{{\cal N}}
\def\nb{{\bf n}}
\def\mb{{\bf M}}
\def\ub{{\bf U}}
\def\ib{{\bf I}}

\def\cnue{c_{\nu_e}}
\def\snue{s_{\nu_e}}

\def\cnumu{c_{\nu_\mu}}
\def\snumu{s_{\nu_\mu}}

\def\snutau{s_{\nu_\tau}}


\vsize=8.5truein
\hsize=6.truein
\hoffset=+4truemm
\baselineskip 12truept

\pageno=0
\rightline{FTUV/93-47}
\medskip
\rightline{UM-TH-93-28}
\medskip
\rightline{CERN-TH.7150/94}
\medskip
\rightline{hep-ph/9402224}\par\noindent
\bs\bs\bs
\centerline{{\bf LIMITS ON NEUTRINO MIXING WITH NEW HEAVY PARTICLES}}
\bs\bs\bs
\centerline{Enrico Nardi$^{a}$, Esteban Roulet$^{b}$ and Daniele
Tommasini$^{c}$}
\bs\medskip
\centerline{$^{a}$\it Randall Laboratory of Physics, University
of Michigan}
\centerline{\it Ann Arbor, MI 48109--1120, U.S.A.}
\medskip
\centerline{$^{b}$\it Theory Division, CERN, CH-1211, Geneva 23,
Switzerland}
\medskip
\centerline{$^{c}$\it Departament de F\'\i sica Te\`orica, Universitat
de Val\`encia, and I.F.I.C.}
\centerline{\it 46100 Burjassot, Valencia, Spain}
\bs
\bs
\bs
\centerline  { {ABSTRACT}}
\medskip
\hskip.3truein
\vbox{
\noindent
\hsize=5.truein
\baselineskip 12truept
We study the effects induced by new neutral fermions below their mass
threshold, due to their possible mixing with the standard neutrinos.
We use as experimental constraints the recent results on lepton
universality, together with the measurement of the $\mu$ decay rate
and the updated LEP data.
In particular, the inclusion in our data set of the most recent
determinations of the $\tau$ branching fractions, mass and lifetime
implies that a previous indication of a non-vanishing mixing for
$\nu_\tau$ is no longer present.
We obtain new stringent limits on the
mixing parameters between $\nu_e$, $\nu_\mu$, $\nu_\tau$
and heavy neutral states of different weak isospin.
If no assumption on the type of neutrinos involved in
the mixing is made, we find $\snue^2<0.0071$, $\snumu^2<0.0014$ and
$\snutau^2<0.033$.

\vskip 2.5truecm
}      

\vskip 1.5truecm
\leftline{January 1994}
\centerline{to appear in Phys. Lett. B}
\vfill
\noindent
--------------------------------------------\phantom{-} \hb
\leftline{Electronic mail: nardi@umiphys,
vxcern::roulet, tommasin@evalvx}

\par\vfill\eject

\interlinea  
\text
\phantom{\cref{ll}\cref{fit}\cref{fit6}}

\centerline{\bf 1. Introduction}

\bigskip

Several extensions of the Standard Model (SM), such as the Left--Right,
SO(10) and E$_6$ models, predict the existence of new neutral fermions.
LEP results imply that these new particles, if they are not singlets under
the SM group, should be heavier than $\sim M_Z/2$.
Even in the case in which these new particles
are too heavy to be directly produced,
their presence could still affect the physics at low energy,
since in general they will be mixed with the SM neutrinos.
Such a mixing will affect the neutral current
(NC) and charged current (CC) couplings of the
known neutral states,
and hence can be constrained by the precision tests of the SM.
Previous model-independent,
comprehensive analyses were performed in refs. [\ncite{ll}{fit6}],
resulting in particular in limits on the
neutrino mixing angles roughly at the level
of $\leq 10^{-1}$.

Recently, new precise tests of universality from the leptonic decays of the
$\pi$ meson [\ccite{triumf92}{psi92}] and from $\tau$ decays
[\ccite{newtau}{roney}] have become available. New determinations of the
$\tau$ mass [\cite{taumass}] and lifetime [\cite{taulife}]
have also removed the previous $\sim 2\sigma$
discrepancy from the SM predictions
in $\tau$ decays\note{In a previous analysis [\cite{fit}]
this discrepancy appeared to be an indication
of a non-vanishing
mixing between $\nu_\tau$ and new heavy
neutral state with weak-isospin $+{1\over 2}$.}.
In addition, the LEP determination of the invisible width of the $Z$ boson
has also improved.

In this paper we take into account all these results, including also the
experimental data on $\mu$ decay, and
derive new stringent limits on the mixing angles of the known neutrinos
with new neutral fermions heavier than a few GeV.

{}From the theoretical point of view,
these results are particularly important in order to
constrain a class of models in which
large mixing angles of the standard neutrinos with new states
are allowed, keeping at the same time the masses
of the known neutrinos below the laboratory limits.
As we will discuss in Section 2, these models require
an (almost) degenerate mass matrix for the neutral states,
so that vanishing (very small) masses for SM
neutrinos can result in a natural way.

\bigskip

\centerline{\bf 2. Mixing with new heavy neutrinos}

\bigskip

We will describe  all the new independent neutral fermionic degrees of
freedom by left-handed fields arranged in a vector $\nu_\n$, without
distinguishing between neutrinos and antineutrinos
[\ncite{ll}{fit6}]. Thus we introduce a vector
$\nb^{0T}=(\nu^T,\ \nu_\n^T)$ of
left-handed spinors, where $\nu^T\equiv
(\nu_{eL},\nu_{\mu L},\nu_{\tau L})$
are the known neutrinos. The fields appearing in
$\nb^0$ are gauge eigenstates. The general mass term
for the neutral states is
$$
{\cal L}_{\rm mass}=\half \nb^{0T} C {\bf M} \nb^0
,\eq{fitnu1}
$$
where $C$ is the matrix of charge conjugation.
The matrix ${\bf M}$ is symmetric and can be
diagonalized through an ``orthogonal" transformation
$\ub^T \mb\ub=\mb_{diag}$, where $\ub$ is {\it unitary},
$$
\nb^0=\ub \nb,
\qquad\qquad \ub^\dagger\ub=\ib
.\eq{fitnu2}
$$
We can write the vector $\nb$ of the mass eigenstates as
$\nb^T=(n^T,\ n_h^T)$,
where $n$ are the light neutrinos and $n_h$ the heavy ones.
Here we assume that the number of states in the vector $n$ is three,
corresponding to the number of standard neutrinos measured at LEP.
The matrix $\ub$ can then be written in block form as
$$
\ub=\pmatrix{A&G\cr
             F&H\cr}
,\eq{fitnu3}
$$
so that the unitarity conditions read
$A^\dagger A +F^\dagger F = A A^\dagger + G G^\dagger = I$.

We will allow for the general
possibility that different kinds of new neutrinos, with different
$SU(2)$ transformation properties, be present.
Then
the vector $\nu_\n$ can be decomposed in subspaces labelled
by different values of the
weak isospin $t_3$,
$\nu_\n^T =(\nu_{t_3=1/2}^T,\nu_{t_3=0}^T,\nu_{t_3=-1/2}^T,
\nu_{t_3=+1}^T,\nu_{t_3=-1}^T,...)$.
For instance, the subspace labelled by
${t_3=0}$ contains neutrinos from singlets
(or from real triplets, which sometimes
have been considered in the
literature [\cite{trip}], or even from larger real multiplets).
The $t_3=1/2$ subspace contains new ordinary neutrinos (i.e. transforming
as the standard doublet neutrinos), and the $t_3=-1/2$
subspace contains exotic neutrinos belonging to right-handed
doublets that can appear in models with mirror and/or vector
doublets of leptons.
All these possibilities are simultaneously present, for example,
in $E_6$ models.
The above formalism allows also for other
possibilities, e.g. $t_3=\pm 1$, corresponding to neutrinos appearing
in $SU(2)_L$ complex triplets. Accordingly the matrices $F$ and $H$
can also be decomposed in submatrices corresponding to different subspaces,
e.g. $F^T=(F_{t_3=1/2}^T,F_{t_3=0}^T,...)$.

The charged current
$J^\mu_W=\bar \nu\gamma^\mu e_L$,
projected onto the subspace of the
light states, reads
$$
J^\mu_W=\bar n\gamma^\mu A^\dagger e_L
.\eq{fitnu4}
$$
In general, whenever the neutrinos mix with neutral states in
non-singlet representations, we can expect that also the left-handed
charged leptons will be mixed with the new charged states present in
the new multiplets. This would give an additional suppression
factor for the CC coupling in \req{fitnu4}, which would result in
constraints on the $\nu$ mixings stronger than the conservative limits
we will obtain below.
In addition to this, an induced right-handed
current term would be present on the r.h.s in \req{fitnu4} whenever
both the neutrinos and the right-handed charged leptons mix with
exotic states in right-handed doublets. As discussed in detail in Refs.
[\ccite{fit}{fit6}], its effect is of higher order in the
light-heavy mixings, and thus can be neglected in the present
analysis. This allows us to restrict the set of new physics parameters
that describe the mixing effects  in the CC observables to the
mixings of the neutrinos, since the corresponding constraints will
still hold in the presence of non-zero mixings for charged leptons.

Denoting with ${\bf T_3}$ the generator corresponding to the third
component of weak isospin,
the NC coupled to the $Z$ boson is $J^\mu_Z=
\bar \nb^0\gamma^\mu {\bf T_3} \nb^0=
\bar \nb\gamma^\mu \ub^\dagger {\bf T_3} \ub \nb$.
When projected onto the light states, this gives
$$
J^\mu_{Z}=\bar n\gamma^\mu \left[\half A^\dagger A + F^\dagger T_3 F
\right] n .\eq{fitnu5}$$
The matrix $T_3$ is diagonal, with entries given
by the values of the weak isospin on the corresponding components
of ${\bf n}^0$.

In processes occurring below
the mass threshold for the production of
the heavy states, the standard gauge
eigenstate $\nu_a$ ($a=e,\mu,\tau$) is effectively replaced by its
(normalized) projection $\vert \nu_a^{light}\rangle$
onto the subspace of the light neutrinos $\vert n_i\rangle$
($i=1,2,3$),
$$
\vert \nu_a^{light}\rangle \equiv {1\over
c_{\nu_a}}\sum_{i=1}^{3} A^\dagger_{ia}
\vert n_i\rangle,
\eq{fitnu6}
$$
where $c_{\nu_a}^2\equiv\cos^2\theta_{\nu_a}\equiv (AA^\dagger)_{aa}$.
The state $\vert \nu_a^{light}\rangle$ has
non-trivial projections on the
subspace of the standard neutrinos $\vert \nu_b\rangle$
as well as on the subspaces
of the new neutrinos $\vert \nu_B^{t_3}\rangle$,
where the superscript refers to the particular value
$t_3$ of the weak isospin. In fact we have
$$
\eqalign{
\sum_b\vert\langle\nu_b\vert\nu_a^{light}\rangle\vert^2=
&{(AA^\dagger)_{aa}^2\over c_{\nu_a}^2}
=c_{\nu_a}^2,\cr
\sum_B\vert\langle\nu_B^{t_3}\vert\nu_a^{light}\rangle\vert^2=
&{(AF_{t_3}^\dagger F_{t_3}A^\dagger)_{aa}\over c_{\nu_a}^2}
\equiv\lambda^a_{t_3} s_{\nu_a}^2,\cr
}\eq{fitnu7}$$
with $s_{\nu_a}^2\equiv 1-c_{\nu_a}^2=\sin^2\theta_{\nu_a}$, and $
\sum_{t_3}\lambda^a_{t_3}=1$ from unitarity.
The parameter $\theta_{\nu_a}$ measures the total amount of mixing
of the known state of flavour $a=e,\mu,\tau$ with the new states,
while
$\lambda^a_{t_3}$ gives the relative weight of the
particular mixing involving new states of weak isospin $t_3$.
These sets of parameters are sufficient to describe the
effects of the light-heavy mixing in all the CC and NC processes that
we will discuss.
Let us consider as an example a weak decay involving the
transition $e_a\to n_i$,
where the subscript $a=e,\mu,\tau$ labels the flavour
and $i=1,2,3 $ corresponds to a light-mass eigenstate.
{}From eq. \req{fitnu4}, the change with respect to the SM
decay rate induced by the neutrino mixings is
$$
{1\over \Gamma_{SM}}\sum_i\Gamma (e_a\to n_i)=(A
A^{\dagger})_{aa} =c_{\nu_a}^2
,\eq{fitnu11}
$$
and depends only on the global reduction $c_{\nu_a}$ in the
couplings of the light neutrinos.

Before
concluding this Section we want to comment on the theoretical expectations
for the values of the mixing parameters. In the usual see-saw mechanism for the
generation of small neutrino masses, the light-heavy mixing angle $\theta$
depends on the ratio between the light and heavy mass scales, and
typically we have $\sin^2\theta \sim m/M$.
For $M\gtrsim M_Z/2$, even if we take $m$ as large as
allowed by the laboratory limit on the $\nu_\tau$ mass
$\sim 31 $ MeV [\cite{pdg92}]\note{Such a
large mass would require an unstable $\nu_\tau$
($\tau({\nu_\tau})\lesssim$ 40 s)
not to conflict with the cosmological limits derived from big bang
nucleosynthesis  [\cite{nucleonu}].},
we get
$\sin^2\theta\lesssim 10^{-3}$, that is more than one order of
magnitude smaller than the constraints that we will derive from the
experimental data in Section 4.

In general it is hard to obtain larger mixing
of the standard neutrinos with particles heavier than $\sim M_Z/2$
without conflicting with the laboratory limits on the $\nu$ masses
[\cite{pdg92}], unless the matrix $\mb$ is (almost) degenerate.
This can occur if (at
least) two kinds of new states are present. For instance, let us assume
that a pair $N$, $N^c$ of new neutrinos exists,
with the lepton number assignments
$L(N)=-L(N^c)=L(\nu)=1$, and that $L$ is conserved.
Then, in the basis $\nb^T=(\nu^T, N^T, N^{cT})$, the
mass matrix is
$$
\mb=\pmatrix{0 & 0 & M_{\nu N^c}\cr
             0 & 0 & M_{N N^c}\cr
M_{\nu N^c}& M_{NN^c} & 0\cr},
\eq{fitnu8}
$$
which is degenerate, ensuring that three eigenstates form
massless Weyl neutrinos. This is due to the fact that
while the heavy states form Dirac neutrinos, the light states remain
with no chirality partner and hence massless.
However, small $L$-violating Majorana mass terms
for the states $\nu$ and $N$ could also be allowed,
and could be relevant for explaining
the solar neutrino deficit via neutrino oscillations.

Mass matrices of the form \req{fitnu8}
can arise for example in generalized $E_6$ models
[\ccite{mohapatra-valle86} {unconventional}],
as well as in models predicting other kinds of vector multiplets
(singlets, triplets, \dots) or new  mirror multiplets of
leptons with neutral components $N$, $N^c$.

Clearly, if in \req{fitnu8} $M_{\nu N^c}\sim M_{N N^c}$, the
mixing angle between $\nu$ and $N$
can be arbitrarily large.
For example if the new neutrinos $N$ are {\it ordinary}, then
$M_{\nu N^c}$ and $M_{N N^c}$ could be generated by
vacuum expectation values of Higgs fields
transforming in the same way under $SU(2)$
so that the $\nu$-$N$ mixing could be naturally close to maximal.

A different possibility would be to have light singlets, whose mass is
not constrained by LEP and hence the see-saw suppression is less
severe. Besides the constraints that we obtain below, which hold for
$m_{\nu_s}\gtrsim {\rm GeV}$, significant bounds also exist for lighter
singlets ($m_{\nu_s}\ll {\rm GeV}$) coming from primordial
nucleosynthesis, since their mixing with active states can bring them,
via oscillations, into equilibrium [\cite{nsbounds}].

\bigskip

\centerline{\bf 3. Experimental constraints}

\bigskip

All the precise laboratory experiments are in good agreement
with the predictions of the SM, which depend on a set of
fundamental input parameters.
Here we choose  the QED coupling
constant $\alpha$ measured at $q^2=0$, the mass of the $Z$ boson
$M_Z=91.187\pm0.007$ GeV [\cite{lep93}] and
the Fermi constant $G_F$.
All the relevant one-loop corrections have been taken into account
in the numerical analysis.

The values of $\alpha$ and $M_Z$ as extracted from
experiments are not affected by the mixings.
In contrast, the effective $\mu$-decay constant
$G_\mu = 1.16637(2) \times 10^{-5}\ {\rm GeV}^{-2}$
is related to the fundamental coupling $G_F$ through the neutrino mixing
angles [\ncite{ll}{fit6}],
$$
G_\mu = G_F c_{\nu_e}c_{\nu_\mu}.
\eq{fitnu9}$$
As a consequence, all the observables that depend
on the strength of the weak interactions $G_F$ will be affected by
the mixing angles $\theta_{\nu_e},\theta_{\nu_\mu}$.
This is the case, for instance, for the
Cabibbo--Kobayashi--Maskawa (CKM) matrix elements.

For our analysis we have used the CC constraints on
lepton universality and on CKM unitarity,
as well as the NC constraints from the LEP measurements
at the $Z$ peak.

\medskip
\no
{\it Lepton universality}

The ratios $g_e/g_\mu$ and $g_\tau/g_\mu$ of the leptonic couplings
to the $W$ boson, which in the SM are predicted to be unity
(universality) are extracted from weak decays.
{}From eq. \req{fitnu11}
we get
$$
\left(g_a\over g_\mu\right)^2 = {c_{\nu_a}^2 \over \cnumu^2 }
,\qquad
a=e,\tau.
\eq{fitnu12}
$$

The best test of $e$-$\mu$ universality comes
from $\pi\to e\nu$ compared to $\pi\to \mu\nu$. The ratio
$
R_\pi\equiv {\Gamma(\pi\to e\nu+\pi\to e\nu\gamma) /
\Gamma(\pi\to \mu\nu+\pi\to \mu\nu\gamma)}
$
has been recently measured with great accuracy at TRIUMF,
$R_\pi=[1.2265\pm0.0034({\rm stat})\pm0.0044({\rm syst})]
\times10^{-4}$ [\cite{triumf92}], and at the PSI,
$R_\pi=[1.2346\pm0.0035({\rm stat})\pm0.0036({\rm syst})]
\times10^{-4}$ [\cite{psi92}].
Combining systematic and statistical errors in quadrature,
the average of the two experiments is
$R_\pi=[1.2310\pm0.0037]\times10^{-4}$, to be compared with the theoretical
SM prediction $R_\pi^{SM}=[1.2352 \pm 0.0005]\times 10^{-4}$
[\cite{marciano-sirlin93}].
Thus we obtain
$
\left(g_e/ g_\mu\right)^2= {R_\pi/ R_\pi^{SM}}=0.9966\pm0.0030
$.

The best tests of $\mu$-$\tau$ universality come from
the $\tau$ leptonic decays compared to $\mu$ decay. The new world
average, including LEP data, gives
$(g_\tau/g_\mu)^2=0.989\pm0.016$ [\cite{roney}].
This value takes into account also the
recent improvements on the determinations of the $\tau$ mass
[\cite{taumass}] and lifetime [\cite{taulife}],
and is no longer in conflict with the SM prediction of unity.
A second test comes from $\tau\to\pi(K)\nu_\tau$,
which gives $(g_\tau/g_\mu)^2=1.051\pm0.029$ [\cite{roney}];
this is almost $2\sigma$ off the SM,
and hardly compatible with the above determination from $\tau$ decays.
In our analysis, we have used the first, more accurate determination
of $(g_\tau/g_\mu)^2$, which is also theoretically clearer.
However, we will also give the results obtained including
both these constraints at the same time.

\medskip
\noindent{\it CKM unitarity}

The {\it observed} CKM matrix elements
$V_{ud}$ and $V_{us}$ are obtained by dividing by $G_\mu$ the measured
vector coupling in $\beta$ decay and in $K_{e3}$ and hyperon decays,
respectively. These processes are weak decays involving
a vertex with $\nu_e$, so that a factor $\cnue$ is present.
Then the effect of the mixing is to modify $V_{ui}$, $i=d,s$, by a factor
$\cnue G_F/G_\mu$
[\ccite{ll}{fit}].
The value of $|V_{ub}|$, obtained from the analysis of semileptonic
$B$ decays, is negligibly small for our purposes. Then the unitarity
constraint for the first row of the CKM matrix in the presence of
neutrino mixing becomes [\ccite{ll}{fit}]
$$
\sum_{i=1}^3|V_{ui}|^2 = \left( {G_F\over G_\mu}\cnue\right)^2.
\eq{fitnu10}
$$
Due to eq. \req{fitnu9}, this depends only on $\snumu$. This theoretical
expression has to be compared with the result
extracted from experiments [\cite{sirlin93}]:
$\sum_{i=1}^3|V_{ui}|^2 = 0.9992\pm 0.0014 $.

\medskip
\no
{\it Invisible width of the $Z$ boson}

The invisible decay rate of the $Z$ boson into neutrinos is
proportional to the Fermi constant $G_F$ multiplied by
the sum of the squares of the neutrino NC couplings
in \req{fitnu5}.
Normalizing to the SM contribution, and keeping only the first
order in the light-heavy mixing, we get
$$
{\Gamma_{Z\to {\rm inv}}\over\Gamma_{Z\to {\rm inv}}^{SM}}=
{G_F\over G_\mu}
{{\rm Tr}(A^\dagger A + 2F^\dagger T_3 F)^2\over3}
\simeq {G_F\over G_\mu}
\left[{1-{1\over3}\sum_a \Lambda_a s_{\nu_a}^2 + O(s^4) }\right],
\eq{fitnu13}
$$
where the parameters
$\Lambda^a\equiv2\sum_{t_3}(1-2t_3) \lambda^a_{t_3}$
describe the particular admixture of heavy neutrinos,
assumed heavier than $M_Z/2$, involved in the mixing.

If the light states are mixed only with new ordinary
states, corresponding to $\lambda_{t_3=1/2}=1$ ($\Lambda=0$),
the invisible width is
slightly {\it increased} by the factor $G_F/G_\mu$.
A mixing with singlet
neutrinos ($\lambda_{t_3=0}=1$, $\Lambda=2$) or with neutrinos from exotic
doublets ($\lambda_{t_3=-1/2}=1$, $\Lambda=4$)
will give a {\it reduction} in the effective number of neutrinos.
Neutrinos appearing in different kinds of lepton triplets behave
in different ways. For real triplets the neutral component has
$t_3=0$ and behaves essentially like a singlet. Neutrinos in complex
triplets have  $t_3=\pm1$, so that $\Lambda=-2$ or $\Lambda=6$.

We see that the effects of light-heavy neutrino mixing
can be described by means of the parameters
$\lambda_{t_3}^a$ (appearing
in the three combinations $\Lambda_a$)
and $\theta_{\nu_a}$,
which together define the decomposition \req{fitnu6}
of the vector $\vert \nu_a^{light}\rangle$.


As we discussed in Section 2, in the presence of mixings in the
neutral sector it is natural to expect that also the charged
leptons will be mixed with new heavy partners, implying a possible
violation of universality for their couplings to the $Z$ boson.
We then use, for consistency, a determination of $\Gamma_{Z\to {\rm
inv}}$ from the LEP experiments, which is independent of the assumption
of universality for the charged-lepton couplings.
Let us consider the flavour-dependent
$Z$ line-shape parameters $\Gamma_Z,\sigma^0_h,R_e,R_\mu,R_\tau$
[\cite{lep93}], where
$\Gamma_Z$ is the total width of the $Z$ boson, $\sigma^0_h$ is the
hadronic cross section at the peak, and $R_l\equiv\Gamma_h/\Gamma_l$ are the
ratios of the hadronic to the leptonic ($l=e,\mu,\tau$) partial widths.
We can define
$\Gamma_{Z\to {\rm inv}}=\Gamma_Z-\Gamma_h-\sum_{l=e,\mu,\tau}\Gamma_l$.
Then from the relation
$\sigma^0_h={12\pi\over M_Z^2}{\Gamma_h\Gamma_e\over \Gamma_Z^2}$
we get
$$
\Gamma_{Z\to {\rm inv}}=\Gamma_Z\left[1-\sqrt{{M_Z^2\over 12\pi}R_e\sigma^0_h}
\left(1+\sum_l{1\over R_l}\right)\right] .\eq{fitnu14}$$
Using the experimental determinations [\cite{lep93}] for the set
$\Gamma_Z,\sigma^0_h,R_e,R_\mu,R_\tau$ and its correlation matrix, we
obtain $\Gamma_{Z\to {\rm inv}}=(498.6\pm5.2)$ MeV,
slightly different from the value obtained in ref. [\cite{lep93}]
with the assumption of universality.
Taking into account the SM prediction
$\Gamma_{Z\to {\rm inv}}=3\,({G_\mu M_Z^3/12\sqrt2\pi})\rho=
(497.6\ {\rm MeV})\rho$, with
$\rho\simeq1+{3\over8\sqrt2\pi^2}G_\mu m_t^2$, we finally obtain
${\Gamma_{Z\to {\rm inv}}/\Gamma_{Z\to {\rm inv}}^{SM}}=
(1.002\pm0.010)/\rho$. The dependence on $\rho$ will give a correlation
between the top quark mass $m_t$ and the neutrino
mixing angles (mainly $\snutau$, which is less constrained by other
measurements).

Besides the constraints arising from the invisible width, eq. (13),
obtained under the assumption $M>M_Z/2$, an important signature at the
$Z$ peak of fermionic mixing is the single production of a heavy state
(provided it is lighter than $M_Z$) in association with a light one.
In this case,
visible signals due to the decay of the heavy neutrino
could be present [\cite{gronau}, \cite{dittmarea90}],
since the mixing will also imply that
the heavy neutrino can decay through the channels $n_h \to \nu + Z^*
\to \ell^+\ell^-,q\bar q$ and $n_h \to \ell + W^* \to \ell
\nu_\ell^-,q q^\prime$. A search for signals of these decays has been
performed at LEP for heavy singlets in the mass range 3 GeV $< M< M_Z$
[\cite{opal-single}, \cite{l3-single}]. For 3 GeV $< M < 50$ GeV the
resulting limits, $s^2_{\nu_a}\lesssim 10^{-4}$, are very stringent,
and they worsen for $M$ approaching $M_Z$ due to
phase space suppression of the production rate [\cite{l3-single}].
This mechanism provides then a tool to search for tiny mixing angles
with states lighter than $M_Z$, a search that is complementary to the
ones discussed in this paper.
\bigskip
\centerline{\bf 4. Results}
\bigskip

We have collected all the theoretical predictions and the experimental
results for the electroweak observables in a $\chi^2$ function.
We have verified that the zero mixing case
($s_{\nu_a}=0$ for $a=e,\mu,\tau$) lies inside
the 90\% confidence level region, so that there are no evidences for
neutrino mixing with new particles.
The former $2\sigma$ disagreement of the data from $\tau$ decays
has disappeared, owing to the new experimental
results discussed in Section 3. As a result, also the signal
of a possible non-zero mixing
of the $\tau$ neutrinos with new ordinary
states, which was found in a previous analysis [\cite{fit}],
is no longer present.
Since no signals of non-zero mixings are found, all
the results of our fit are presented as (90\% c.l.)
upper bounds on the mixing angles.

For the limits on the mixings $\snue$ and $\snumu$ we obtain
$$
\snue^2<0.0071,\qquad\qquad\snumu^2<0.0014,
\eq{fitnu15}
$$
almost independently of $\Lambda_e$ and $\Lambda_\mu$.
This is due to the fact that the main sources of constraints are CC
processes, namely $\mu$ decay and the lepton universality test on
$(g_e/g_\mu)^2$ for $\snue^2$, and the CKM unitarity test
(depending on the $\mu$-decay constant) for $\snumu^2$.
The constraints from the LEP data are much less stringent than the
previous ones, and for this reason
the limits \req{fitnu15} will not change significantly,
even if the new neutral states involved in
the mixing were lighter than $M_{Z}$.
However they are still assumed to be
heavier than the production energy thresholds in the
relevant processes used for the experimental determinations,
that is $m_\pi$ and $m_\Lambda$ respectively.

\midinsert
{
$$
\vbox{\hsize= 13.5truecm
\baselineskip=12pt
\noindent TABLE I. Value of the 90\% c.l. upper bound on $\snutau$,
corresponding to a set of values of the parameter $\Lambda_\tau$
describing the type of the new heavy particles involved in the mixing
with $\nu_\tau$.
The top mass has been fixed to $m_t=150$ GeV.
The first line gives the conservative limits,
obtained using the set of constraints including the data on
$\mu$ and $\tau$ leptonic decays and on the invisible width
of the $Z$ boson.
The tighter bounds given in the second line
are less conservative, since they
have been obtained by including in the data set
also the constraint from $\tau\to\pi(K)\nu_\tau$.
\vskip -.8truecm}
$$

$$
\vbox{\hsize= 13.5truecm
\def\f{\phantom{m}}
\baselineskip=14pt
\halign{ &\strut#&\qquad#\f\hfil&\hfil\f#\f\hfil\cr
\tabrul2
& & $\Lambda_\tau=0$ & $\Lambda_\tau=2$ & $\Lambda_\tau=4$ &
$\Lambda_\tau=6$ & $\Lambda_\tau=-2$ \cr
\tabrule
& Conservative bound & 0.033  & 0.024 & 0.015 & 0.010 & 0.019 \cr
\tabrule
& Using $\tau\to\pi(K)\nu_\tau$ & 0.020  & 0.017 & 0.012 &
0.0089 & 0.013 \cr
\tabrul2
}}$$
\bs
}
\endinsert
\interlinea

In contrast, the limits on $\snutau$ do depend on the value of
$\Lambda_\tau$, that is on the weak isospin of the new neutrinos
involved in the mixing. The different limits are given in table I. The
conservative bounds are obtained using only the determination of the
universality ratio from $\tau$ decays and the results from LEP. In the
second line we give the more stringent, but less conservative, bounds
obtained by including in the data set the constraint from
$\tau\to\pi(K)\nu_\tau$.
The result for $\Lambda_\tau=0$ coincides also with the bound
that can be set using only the CC constraints, since for this
particular value of $\Lambda_\tau$ the NC couplings are not affected
by $\snutau$. This bound,
$$
\snutau^2<0.033 \quad(\hbox{$0.020$ including
$\tau\to\pi(K)\nu_\tau$}),
\eq{fitnu16}
$$
is valid for any $\Lambda_\tau$ provided that the new particles
involved in the mixing are heavier than $m_\tau$, and improves by a
factor 3 (5) the previous result ($\snutau^2<0.098$ for
$\Lambda_\tau=0$ [\cite{fit}]). For $\Lambda_\tau\ne0$ the bounds
are more stringent, due to the effectiveness of the additional
constraint from the $Z$ invisible width. Assuming that the new
particles are heavier than $M_Z$, and taking the value $m_t=150$ GeV
for the top-quark mass, this constraint alone sets a limit
$\snutau^2<0.060/\Lambda_\tau$ for $\Lambda_\tau>0$
($\snutau^2<0.040/\vert\Lambda_\tau\vert$ for $\Lambda_\tau<0$).
For $\vert\Lambda_\tau\vert\gtrsim1$ this is comparable to the effect
of the CC constraint, and the combined result is given in the table.
For large $\vert\Lambda\vert$ the constraint from LEP data is the most
stringent though somewhat less general, since it holds only under the
assumption that the new particles involved in the mixing are heavier
than $M_Z$.
The bounds change slightly (at most by $\sim20$\% of their value)
for different choices of $m_t$ in the range
110 GeV $<m_t<$ 200 GeV. In general, since
$\Gamma_{Z\to inv}\sim (1-{1\over3}\Lambda_\tau\snutau^2)
(1+{3\over8\sqrt2\pi^2}G_\mu m_t^2)$ the limit is
weakened either for greater $m_t$
if $\Lambda_\tau>0$, or for smaller $m_t$ when $\Lambda_\tau<0$.

As we have already mentioned, models predicting new neutral fermions
often  contain also additional charged fermions as well as new neutral
gauge bosons [\ccite{fit}{fit6}]. Then in some observables the
contributions due to these additional sources of new physics can
compensate in part the effects of the light--heavy neutrino mixing. As
we have discussed, our limits are still reliable in the presence of
non-zero mixings for the charged leptons. However, as is shown in
references [\ccite{fit}{fit6}], the additional effects of new neutral
gauge bosons and especially of non-zero light-heavy  mixings in the
quark sector can relax the bounds on the neutrino mixing angles. The
most important cancellations can affect the bound on $\snumu$, whose
contribution to the CKM unitarity sum can be compensated by a mixing
between the left-handed quarks and new exotic particles
[\ncite{ll}{fit6}].

\bigskip

\centerline{\bf 5. Conclusions}

\bigskip

We have studied the effects induced by new neutral fermions  below the
threshold for their direct production.
The possible mixing of these new particles with the standard neutrinos
would affect the NC and CC processes for the light states as discussed in
Section 2.
The CC experimental test on lepton universality and $\mu$ decay,
as well as the NC data on
the $Z$ boson invisible width, have been used to constrain the
light--heavy neutrino mixing
angles. We have found no evidence for this kind of new physics
effects,
and the new limits we have obtained on the mixing parameters
improve the results of previous analyses [\ncite{ll}{fit6}].
We have also generalized our discussion to include
new states of arbitrary weak isospin.
Our main results are summarized by eqs. \req{fitnu15} and \req{fitnu16}.
These limits are general in the sense that they
hold independently of the weak isospin of
the new states involved in the mixing.
The underlying assumption that the new states are heavier than
$m_\pi$, $m_\Lambda$, $m_\tau$, respectively, for the neutrino mixings
of the first, second and third generation, is non-trivial only for the
mixing with singlet neutrinos (or more in general, neutrinos
with third component of weak isospin $t_3=0$).
In the particular case of $s_{\nu_\tau}$, the limits can be more
stringent depending on the specific value assumed for the weak
isospin of the heavy states. The
constraints from the LEP measurement of the invisible width is
very important in this case.
A list of limits that correspond to different possible values of
the weak-isospin for the new states is given in table I.

Our bounds are of little use in the framework of
see-saw models, which in fact predict light-heavy mixing angles
much smaller than the experimental limits.
However, different models exist in which
the mixing angles between the known and the
new  heavy neutral states can be naturally large,
and they are directly constrained by our results.
For example we have shown that this is the case for
a class of models that predict an (almost)
degenerate mass matrix in the neutral sector, since
in this case large mixing angles are naturally consistent
with vanishing (small) masses for the standard neutrinos.

\bs

\no We would like to thank E. Akhmedov and J. Valle for useful discussions.

\vfill\eject


\def\bi{\biblitem}
\def\hbup{\hfill\break\baselineskip 12pt}
\null
\baselineskip 8pt
\centerline{\bf References}
\vskip .6truecm
\bi{triumf92}
D.I. Britton et al., \prl 68 (1992) 3000. \par
\bi{psi92}
G. Czapek et al., \prl 70 (1992) 17. \par
\bi{newtau} 
N. Colino (L3), in
{\it Proc. of the Second Workshop on Tau Lepton Physics}
ed. K.K. Gan (1992), (World Scientific, Singapore, 1993);\hbup
S. Snow  (ALEPH), {\it ibidem};\hbup
J. Hobbs (OPAL), {\it ibidem};\hbup
P. Vaz (DELPHI), {\it ibidem};\hbup
N. Mistry (CLEO II), {\it ibidem}.\par
\bi{roney}
J.M. Roney, in Ref. [\cite{newtau}].\par
\bi{taumass}
CLEO collaboration,  R. Ballest et al., preprint CLNS-93-1194;\hbup
BEPC collaboration, J. Z. Bai et al., \prl{69} (1992) 3021;\hbup
D. Britton, in Ref. [\cite{newtau}];
H. Marsiske, {\it ibidem}. \par
\bi{taulife}
ALEPH collaboration,  \plb 279 (1992) 411; \plb 297 (1992) 432;\hbup
OPAL collaboration,  \zpc 59 (1993) 183;\hbup
DELPHI collaboration,  preprint CERN-PPE/93-12 (1993);\hbup
M. Uzerman (L3 collaboration), in Ref. [\cite{newtau}];
W. Trischuk, {\it ibidem}; \hbup
CLEO collaboration,  M. Battle et al., \plb 291 (1992) 488;\par
\bi{trip}
B.W. Lee, \prd{6} (1972) 1188; \hbup
J. Prentki and B. Zumino, \npb{47} (1972) 99; \hbup
P. Salati, \plb{253} (1991) 173. \par
\bi{dittmarea90}
M. Dittmar, A Santamaria, M.C. Gonzalez Garcia and J.W.F. Valle,
\npb 332 (1990) 1. \par
\bi{mohapatra-valle86}
R.N. Mohapatra and J.W.F. Valle, \prd 34 (1986) 1642;
J. Bernab\'eu et al., \plb 187 (1987) 303.\par
\bi{unconventional}
E. Nardi, \prd 48 (1993) 3277.\par
\bi{lep93}
The LEP Collaborations and The LEP Electroweak Working Group, preprint
CERN-PPE/93--157 (1993). \par
\bi{marciano-sirlin93}
W.J. Marciano and A. Sirlin, preprint NYU--TH--93--09--03 (1993). \par
\bi{sirlin93}
A. Sirlin, in {\it Precision Test of the Standard Electroweak Model}
(World Scientific, Singapore, 1993), ed. P. Langacker. \par
\bi{nucleonu}
E. Kolb, M.S. Turner, A. Chakravorty and D.N. Schramm, \prl{67} (1991)
553;  A.D. Dolgov and I.Z. Rothstein, \prl{71} (1993) 476; \hbup
M. Kawasaki et al., Report OSU-TA-5/93 and UPR-0562T (1993).\par
\biblitem{pdg90}
Rev. of Part. Prop., \plb 239 (1990) 1. \par
\biblitem{pdg92}
Particle Data Group, \prd 45 (1992) 1. \par
\biblitem{missingtau}
T.N.Truong, \prd  30 (1984) 1509; \hbup
F.J. Gilman and S.H. Rhie, \prd  31 (1985) 1066; \hbup
F.J. Gilman, \prd  35 (1987) 3541. \par

\biblitem{taumix}
Xue-Quian Li and Tao Zhi-jian, \prd 43 (1991) 3691. \par


\bi{rhoth}
M. Veltman, Nucl.Phys. B 123 (1977) 89; \hbup
M.B. Einhorn, D.R.T. Jones, M. Veltman, Nucl.Phys. B 191 (1981) 146.\par
\bi{heavyloops}
M.E. Peskin and T. Takeuchi \prl 65 (1990) 964; \hbup
D.C. Kennedy  and P. Langacker \prl 65 (1990) 2967; E: ibid 66
(1991) 395; UPR--0467T (Mar. 1991); \hbup
W.J. Marciano and J.L. Rosner, Phys. Rev. Lett. 65 (1990) 2963; \hbup
G. Altarelli and R. Barbieri, \plb 253 (1991) 161. \par
%
\bi{vudth}
W. Jaus and G. Rasche, \prd  41 (1990) 166; \hbup
D. H. Wilkinson, TRI--PP--90--44 (1990);
\plb  241 (1990) 317. \par
\bi{ckmth}
A. Sirlin, CU--TP--505 (1990); \hbup
D. G. Hitlin, CALT--68--1722 (1991). \par
\bi{marciano}
W. J. Marciano, BNL--45999 (1991). \par
\bi{jegerlehner}
F. Jegerlehner, PSI--PR--89--23 (1989); PSI--PR--91--08;16 (1991).\par

\bi{leprad}
M. Consoli and W. Hollik, `Z physics at LEP 1' vol. 1,
     eds. G. Altarelli et al., CERN 89--08; \hbup
G. Burgers and F. Jegerlehner, ibidem; \hbup
W. Hollik, CERN--TH.5661/90 (FEB. 1990); \hbup
G. Burgers and W. Hollik, in `Polarization at LEP' vol. 1,
eds. G. Alexander et al., CERN 88--06; \hbup
B. W. Lynn, M. E. Peskin and R. G. Stuart, in `Physics at LEP' vol. 1,
eds. J. Ellis and R. Peccei CERN 86-02 (1986); \hbup
D.C. Kennedy and B.W. Lynn, Nucl. Phys. B 322 (1989) 1; \hbup
J.G. Im, D.C. Kennedy, B.W. Lynn and R.G. Stuart, Nucl. Phys. B
     321 (1989) 83; \hbup
J.L. Rosner, EFI 90--18 (June 1990). \par
\bi{afbth}
D. Bardin et al.,  \plb  229 (1989) 405; \hbup
M. B\"ohm et al., CERN--TH--5536/89 (1989); \hbup
Z. Was and S. Jadach, \prd  41 (1990) 1425; \hbup
D. Bardin et al., \npb  351 (1991) 1;  \hbup
D. Bardin et al.,  \plb  255 (1991) 290; \hbup
S. Jadach, M. Skrzypek and B. F. L. Ward, \plb  257 (1991) 173; \hbup
S. Jadach, B. F. L. Ward and Z. Was, \plb  257 (1991) 213; \hbup
A. A. Akhundov, D. Y. Bardin and A. Leike, \plb  261 (1991) 321.  \par
\bi{bvertex}
A.A. Akhundov, D. Bardin and T. Riemann; Nucl. Phys. B 276 (1988) 1;\hbup
F. Diakonos and W. Wetzel, HD--THEP--88-21 (1988); \hbup
W.Beenakker and W.Hollik, Z. Phys. C40 (1988) 141; \hbup
B.W. Lynn and R.G. Stuart, Phys.Lett. 252 B (1990) 676. \par
\bi{apvrad}
W. J. Marciano and A. I. Sanda, \prd 17 (1978) 3055; \hbup
W. J. Marciano and A. Sirlin, \prd  27 (1983) 552; \hbup
W. J. Marciano and A. Sirlin, \prd  29 (1984) 75. \par

\bi{apv-matrix}
S.A. Blundell, W.R. Johnson and J. Sapirstein, \prl 65 (1990) 1411.\par
\bi{nurad}
W. J. Marciano and A. Sirlin, \prd  22 (1980) 2695; \hbup
A. Sirlin and W. J. Marciano, \npb 189 (1981) 442; \hbup
W. J. Marciano and A. Sirlin, \prd  29 (1984) 945; err. 31 (1985) 213; \hbup
S. Sarantakos, A. Sirlin and W. J. Marciano, \npb  217 (1983) 84. \par
\biblitem{costa}
G. Costa et al., \npb  297 (1988) 244.\par
\biblitem{amaldi}
U. Amaldi et al., Phys.Rev. D 36 (1987) 1385. \par
%
\bi{fitop}
J. Ellis and G.L. Fogli, Phys. Lett. B 213  (1988) 526; \hbup
J. Ellis and G.L. Fogli, Phys. Lett. B 232 (1989) 139; \hbup
A. Blondel, CERN--EP/90--10, (January 90);  \hbup
J. Ellis and G.L. Fogli, Phys. Lett. B 249 (1990) 543.\par
\bi{topnew}
J. Ellis and G.F. Fogli, \plb 249 (1990) 543; \hbup
P. Langacker, UPR--0435T (Aug 1990). \par
%
%
\bi{earlymix}
J. Maalampi, K. Mursula and M. Roos, \npb  207 (1982) 233; \hbup
J. Maalampi and K. Mursula, \zpc  16 (1982) 83; \npb  269 (1986) 109; \hbup
K. Enqvist, K. Mursula and M. Roos, \npb  226 (1983) 121; \hbup
K. Enqvist, J. Maalampi and M. Roos, \plb  176 (1986) 396; \hbup
T. Rizzo, \prd  34 (1986) 2076; \prd  34 (1986) 2163; \hbup
P. M. Fishbane, R. E. Norton and M. J. Rivard, \prd  33 (1986) 2632; \hbup
V. Barger, R. J. Phillips and K. Whisnant \prl  57 (1986) 48; \hbup
M. Gronau, C. N. Leung and L. Rosner, \prd  29 (1984) 2539. \par
\bi{champ}
A. de R\'ujula, S. Glashow and U. Sarid; \npb 333 (1990) 173; \hbup
E. Nardi and E. Roulet, \plb  245 (1990) 105.\par
\biblitem{ll}
P. Langacker and D. London, \prd  38 (1988) 886.\par
\biblitem{bot}
E. Nardi and E. Roulet, \plb  248 (1990) 139.\par
\bi{fit}
E. Nardi, E. Roulet and D. Tommasini, \npb 386 (1992) 239. \par
\bi{fit6}
E. Nardi, E. Roulet and D. Tommasini, \prd 46 (1992) 3040. \par
\bi{llm}
P. Langacker, M. Luo and A.K. Mann, UPR-458T (1991).\par
\biblitem{ll2}
P. Langacker and D. London, \prd  38 (1988) 907.\par
\biblitem{ll3}
P. Langacker and D. London, \prd  39 (1989) 266.\par
%
%
\biblitem{zprimo}
R. Gatto et al., in ``$Z$ physics at LEP"
CERN 89--08, vol. 2 (1989) 147;\hbup
B.W. Lynn, F.M. Renard, C. Verzegnassi, Nucl.Phys. B 310 (1988) 237; \hbup
F. Boudjema, F.M. Renard, C. Verzegnassi, Nucl.Phys. B 314 (1988) 301;\hbup
      Phys.Lett. B 202 (1988) 411; Phys.Lett. B 214 (1988) 151;\hbup
G. Altarelli et al., CERN--TH--5626/90; \hbup
F. del Aguila, J.M. Moreno and M. Quir\'os, \prd 40 (1989) 2481;\hbup
E. Nardi, Int. Jour. Mod. Phys. A 6 (1991) 1447. \par
\bi{zprimonew}
J. Layssac, F.M. Renard and C. Verzegnassi, LAPP-TH-290-90-REV.
(1991); \hbup
M.C. Gonzalez Garc\'\i a and J.W.F. Valle; \plb  259 (1991) 365; \hbup
G. Altarelli et al., CERN--TH--6051/91; \hbup
P. Langacker and M. Luo, UPR-0476T (1991); \hbup
F. del Aguila, J.M. Moreno and M. Quir\'os, \prd 41 (1990) 134;
      err. ibid. 42 (1990) 262; \plb 254 (1991) 497. \par
%
%
\bi{rizzo-e6}
for a review, see J.L. Hewett and T.G. Rizzo, Phys. Rep. 183 (1989)
195.\par
\bi{mirror}
J. Maalampi and M. Roos, Phys. Rep. 186 (1990) 53. \par
%
%
%
%
%
\biblitem{wmassth}
A. Sirlin, \prd 22 (1980) 971; \prd 29 (1984) 89; \hbup
W. J. Marciano and A. Sirlin, \prd 29 (1984) 945. \par

\biblitem{wmassexp}
UA2 Collaboration, J. Alitti et al., \plb 241 (1990) 150; \hbup
CDF Collaboration, F. Abe et al., \prl 65 (1990) 2243.\par
%
\bi{apvcs}
M.C. Noecker, B.P. Masterson and C.E. Wieman, \prl 61 (1988) 310.\par
\bi{slac-ed}
C.Y. Prescott et. al., \plb 77 (1978) 347; \plb 84 (1979) 524.\par
\bi{fogli-haidt}
G.L. Fogli and D. Haidt, \zpc  40 (1988) 379. \par
\bi{charm-I-nue}
CHARM Collaboration, J. Dorenbosch et al., \zpc  41 (1989) 567.\par
\bi{charm-II-nue}
CHARM-II Collaboration, D. Geiregat et al., \plb 259 (1991) 499. \par
\bi{bnl-nue}
BNL collaboration, K. Abe et al., \prl  62 (1989) 1709.\par
%
%
\biblitem{lepresults}
E. Fernandez, CERN--PPE/90--151;\hbup
F. Dydak, CERN--PPE/91--14; \hbup
H. Burkhardt and J. Steinberger, CERN--PPE/91--50.\par

\biblitem{alphasexp}
S. Bethke, CERN--PPE/91--36, and references therein. \par

\biblitem{lepgg91}
ALEPH Collaboration, D. Decamp et al., CERN--PPE/91--105; \hbup
DELPHI Collaboration, P. Abreu et al., CERN--PPE/91--95; \hbup
L3 Collaboration, B. Adeva et al., L3 Prep. $\#$ 028 (1991); \hbup
OPAL Collaboration, M. Alexander et al., CERN--PPE/91--81. \par

\biblitem{opalgginv}
OPAL Collaboration, M. Akrawy et al., \zpc 50 (1991) 373. \par

\biblitem{taupolexp}
ALEPH Collaboration, D. Decamp et al.,  CERN--PPE/91--94; \hbup
OPAL Collaboration, G. Alexander et al., CERN--PPE/91--103.\par

\biblitem{nsbounds}
R. Barbieri and A. Dolgov, \plb{237} (1990) 440; \hbup
X. Shi et al., \prd 47 (1993) 3720.

\biblitem{opal-single}
OPAL Collaboration, M.Z. Akrawy et al., \plb 247 (1990) 448. \par

\biblitem{l3-single}
L3 Collaboration, O. Adriani et al., \plb 295 (1992) 371. \par

\biblitem{gronau}
M. Gronau et al., \prd 29 (1984) 2539. \par

\insertbibliografia
\bye